\documentclass[psfig]{aa}
\usepackage{graphicx}
\usepackage{txfonts}
\begin{document}

\title{RAVE spectroscopy of luminous blue variables in the Large Magellanic Cloud}

   \author{U.~Munari\inst{1},
           A.~Siviero\inst{1,5},
           O.~Bienaym\'e\inst{2},
           J.~Binney\inst{3},
           J.~Bland-Hawthorn\inst{4},
           R.~Campbell\inst{5,6},
           K.C.~Freeman\inst{7},
           J.P.~Fulbright\inst{8},
           B.K.~Gibson\inst{9},
           G.~Gilmore\inst{10},
           E.K.~Grebel\inst{11},
           A.~Helmi\inst{12},
           J.F.~Navarro\inst{13},
           Q.A.~Parker\inst{6},
           W.~Reid\inst{6},
           G.M.~Seabroke\inst{10,14},
           A.~Siebert\inst{2,5},
           M.~Steinmetz\inst{5},
           F.G.~Watson\inst{15},
           M.~Williams\inst{5,7},
         R.F.G.~Wyse\inst{8},
           T.~Zwitter\inst{16}}

   \offprints{munari@pd.astro.it}

  \institute{INAF Osservatorio Astronomico di Padova, Asiago, Italy
      \and Observatoire de Strasbourg, Strasbourg, France
      \and Rudolf Pierls Center for Theoretical Physics, University of Oxford, UK
      \and Institute of Astronomy, School of Physics, University of Sydney, Australia
      \and Astrophysikalisches Institut Potsdam, Potsdam, Germany
      \and Macquarie University, Sydney, Australia
      \and RSAA Australian National University, Camberra, Australia
      \and Johns Hopkins University, Baltimore, Maryland, USA
      \and Jeremiah Horrocks Institute for Astrophysics \& Super-computing, University of Central Lancashire, Preston, UK
      \and Institute of Astronomy, University of Cambridge, UK
      \and Astronomisches Rechen-Institut, Zentrum f\"ur Astronomie der Universit\"at Heidelberg, Heidelberg, Germany
      \and Kapteyn Astronomical Institute, University of Groningen, Groningen, The Netherlands
      \and University of Victoria, Victoria, Canada
      \and e2v Centre for Electronic Imaging, Planetary and Space Sciences Research Institute, The Open University, Milton Keynes, UK
      \and Anglo-Australian Observatory, Sydney, Australia
      \and Faculty of Mathematics and Physics, University of Ljubljana, Ljubljana, Slovenia
              }

\date{Received YYY ZZ, 2009; accepted YYY ZZ, 2009}

     \abstract{
              The RAVE spectroscopic survey for galactic structure and
              evolution obtains 8400-8800~\AA\ spectra at 7500 resolving
              power at the UK Schmidt Telescope using the 6dF multi-fiber
              positioner.  More than 300\,000 9$\leq$$I_{\rm C}$$\leq$12 and
              $|b|$$\geq$25$^\circ$ southern stars have been observed to
              date.
              }
              {
	      This paper presents the first intrinsic examination of 
              stellar spectra from the RAVE survey, aimed at evaluating
              their diagnostic potential for peculiar stars and at contributing
              to the general understanding of Luminous Blue Variables (LBVs).
              }
              {
	      We used the multi-epoch spectra for all seven LBVs observed,
              between 2005 and 2008, 
              in the Large Magellanic Cloud (LMC) by the RAVE survey.
              } 
              {
              We demonstrate that RAVE spectra possess significant
  	      diagnostic potential when applied to peculiar stars and, in
  	      particular, LBVs.  The behaviour of the radial velocities for
  	      both emission and absorption lines, and the spectral changes
  	      between outburst and quiescence states are described and found
  	      to agree with evidence gathered at more conventional
  	      wavelengths. The wind outflow signatures and their variability
  	      are investigated, with multi-components detected in
  	      S~Doradus.  Photoionisation modelling of the rich emission line
  	      spectrum of R~127 shows evidence of a massive detached
  	      ionised shell that was ejected during the 1982-2000 outburst.
  	      Surface inhomogeneities in the nuclear-processed material,
  	      brought to the surface by heavy mass loss, could have
  	      been observed in S~Doradus, even if alternative
  	      explanations are possible. We also detect the transition from
  	      quiescence to outburst state in R~71.  Finally, our spectrum
  	      of R~84 offers one of the clearest views of its cool
  	      companion.
              }
	      {}


   \authorrunning{U.Munari et al.}
   \titlerunning{RAVE spectroscopy of LBVs in LMC}

   \maketitle

\section{Introduction}

RAVE (RAdial Velocity Experiment) is an ongoing digital spectroscopic survey
of stars in the magnitude interval 9$ \leq I_{\rm C} \leq$12, distributed
over the whole southern sky at galactic latitudes $|b| \geq$25$^\circ$. 
Spectra are recorded over the 8400-8800~\AA\ range, at a resolving power
$\sim$7500, with the UK Schmidt telescope feeding light to a spectrograph
via the 6-degree Field (6dF) 150 fiber positioner. Via the determination of
radial velocities, chemistries, temperatures, and gravities for a large
number of high-latitude stars, the overarching science driver for the survey
is the investigation of the structure and evolution of the Milky Way. At the
time of writing, RAVE has observed over 300,000 stars, Data Releases 1 and 2
have been published (Steinmetz et al. 2006; Zwitter et al. 2008), the third
is coming (Siebert et al. 2009), and scientific exploitation of the data has
begun (e.g. Smith et al. 2007; Siebert et al. 2008; Seabroke et al. 2008;
Munari et al. 2008; Veltz et al. 2008).

  \begin{figure*}
     \centering
     \includegraphics[height=17.5cm,angle=270]{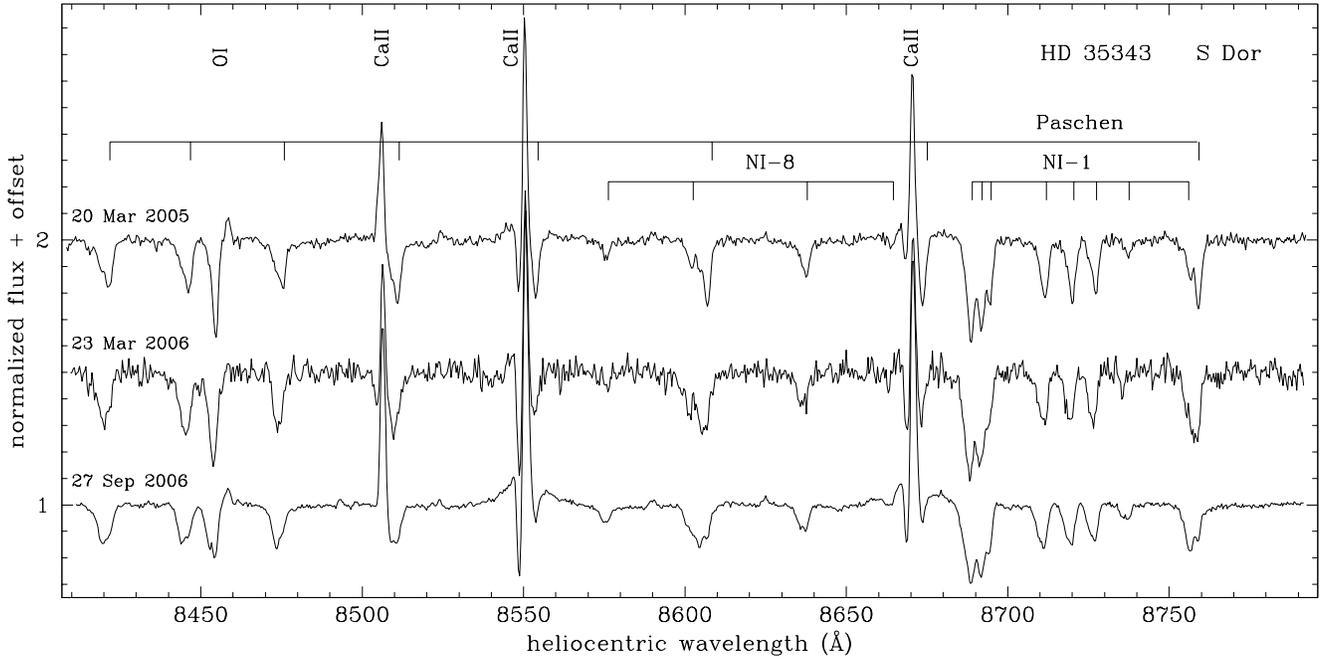}
     \caption{Multi-epoch RAVE spectra of S~Doradus. Paschen and NI 
              (multiplets 1 and 8) absorption lines are indicated 
              by comb markings.}
     \label{fig1}
  \end{figure*}

The essentially unbiased selection of targets guarantees that in addition to
normal stars, peculiar ones are also observed. The type of peculiarities
most easy to recognise with RAVE spectra are the presence of emission lines,
absorption line splitting in double-line binaries, signatures of stellar
winds, and combination spectra. The present paper is the first to explore
the performance of RAVE in relation to the physics of peculiar stars, in
particular, the Luminous Blue Variables (LBVs) of the Large Magellanic
Cloud (LMC).  As part of its tiling strategy, RAVE observed the region of
sky occupied by the LMC in March 2005, March 2006, September 2006, and
February 2008. As LBVs are intrinsically very luminous, and those belonging
to the LMC are brighter than the $I_{\rm C}$$\leq$12 survey limit (indeed,
many possess an HD identifier), they naturally enter the RAVE target set. A
list of the LBVs observed by RAVE in the LMC is given in Table~1.

LBVs were first identified in M31 and M33 by Hubble \& Sandage (1953) as
extremely luminous hot stars that underwent irregular photometric
variability of modest amplitude (a few mag) over timescales of years. Their
bolometric magnitude generally exceeds $-$9.5, corresponding to a luminosity
of the order $\sim$10$^6$~L$_\odot$, close to their Eddington Limit. During
quiescence, the LBVs obey a tight temperature-luminosity relation,
calibrated as $\log L / L_\odot = 1.37 \log T_{\rm eff} - 0.03$ by
van~Genderen (2001). The existence of such a relation, their extreme
intrinsic brightness and easy identification (the latter provided by their
variability), make LBVs important players in the cosmic distance ladder game
(e.g. Davidson et al. 1989). The LBVs are sometimes also called
Hubble-Sandage variables or S~Dor variables from the archetype of the class,
and are believed to be the immediate progenitor of Wolf-Rayet stars of the
classic N-type (nitrogen rich and hydrogen poor). The LBV phase is not
particularly long-lasting ($\sim$0.025~Myrs) and, together with the paucity
of massive stars, accounts for the relative rarity of LBVs (Crowther 2007).
The number of recognised LBVs differs from author-to-author, ranging from
20 to 50 throughout the entire Local Group (e.g. van~Genderen 2001; Weis
2003). The most famous LBVs include $\eta$~Car and P~Cyg, in the Galaxy, and
S~Dor in the LMC.

  \begin{table}
    \caption{List of the LBVs observed by RAVE in the LMC, their date of
	     observation
             and optical brightness at that time, and the radial velocities 
	     of the associated emission and absorption lines.}
     \centering
     \includegraphics[width=8.8cm]{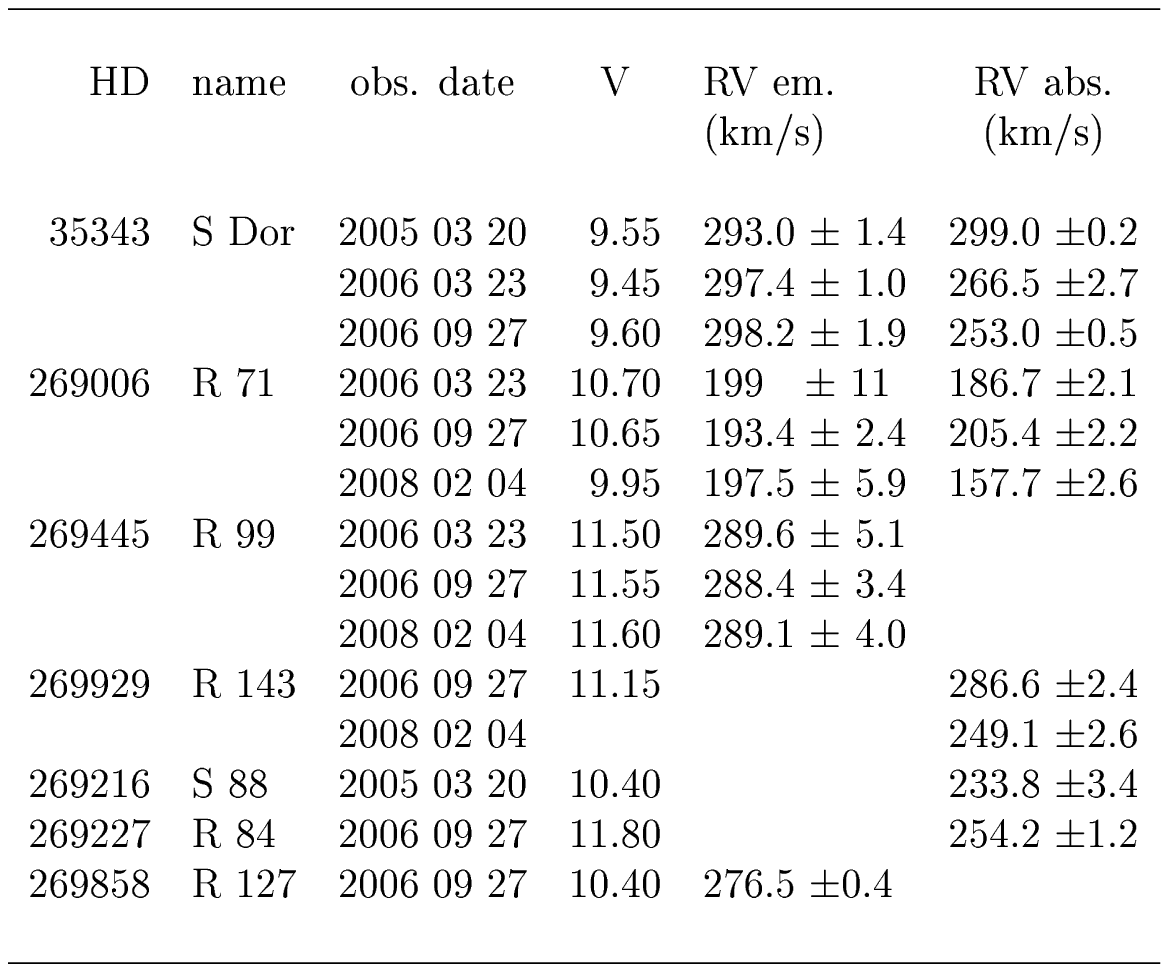}
     \label{tab1}
  \end{table}

  \begin{figure*}
     \centering
     \includegraphics[height=17.5cm,angle=270]{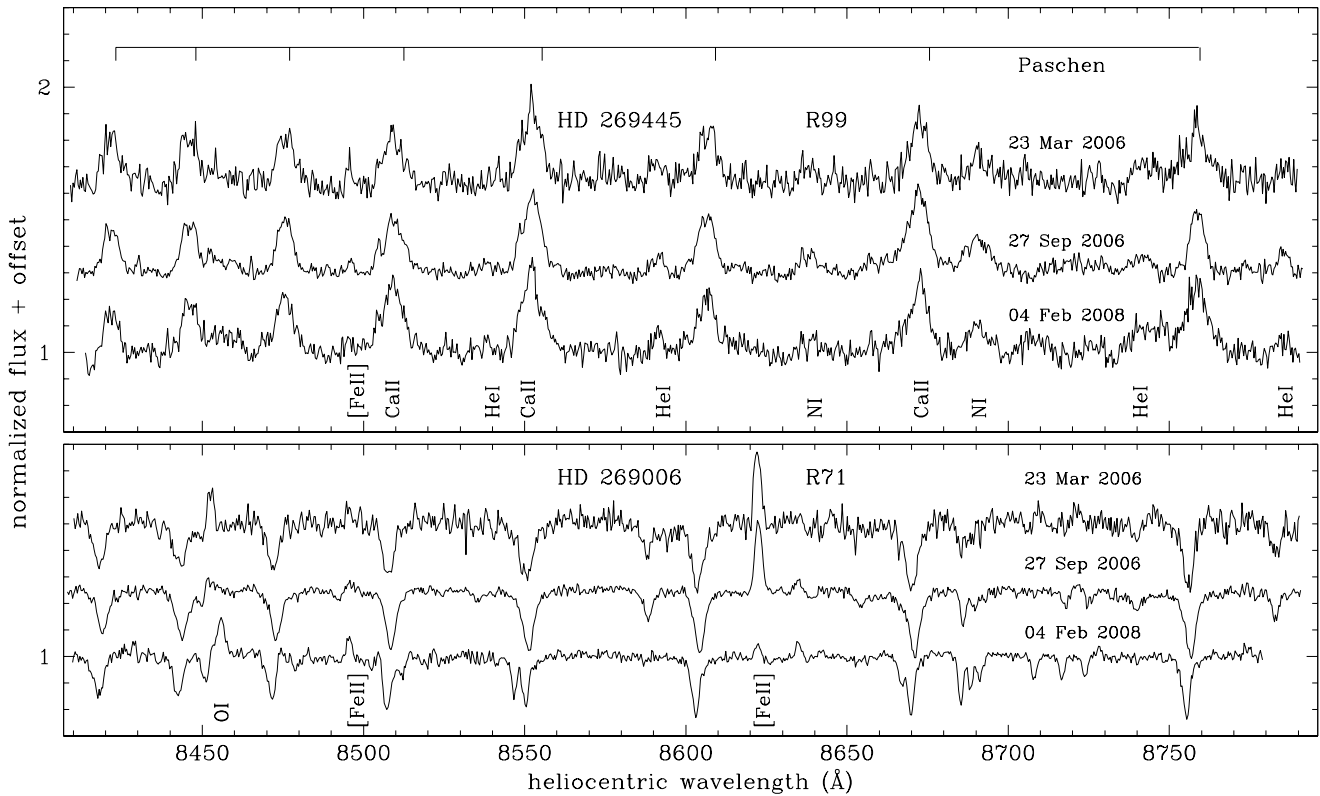}
     \caption{Multi-epoch RAVE spectra of R~99 and R~71.}
     \label{fig2}
  \end{figure*}

After the main sequence phase, the most massive stars do not evolve all the
way to the reddest part of the HR diagram but instead enter a phase of very
high mass loss (10$^{-4}$$-$10$^{-5}$ M$_\odot$~yr$^{-1}$) and reverse their
evolution back toward hotter effective temperatures, becoming LBVs. The
position of this turning point depends upon the star's luminosity and sets
what is called the {\it Humphreys-Davidson Limit} (cf. Humphreys \& Davidson
1994). The LBVs lay to the hotter side of this $T_{\rm eff}$ limit, which
may be crossed during eruptive events of the most massive objects. The
evolutionary sequence leading to the formation of LBVs depends on the mass
of the progenitor. For stars more massive than 75~M$_\odot$ it is O
$\rightarrow$ WN(H-rich) $\rightarrow$ LBV $\rightarrow$ WN(H-poor)
$\rightarrow$ WC $\rightarrow$ SNIc, while for those of initial mass
40$\leq$M$\leq$75~M$_\odot$ it is O $\rightarrow$ LBV $\rightarrow$
WN(H-poor) $\rightarrow$ WC $\rightarrow$ SNIc (cf. Crowther 2007, and
references therein). Direct spectroscopic evidence for LBV progenitors
transitioning to supernovae is being obtained (eg. SN\,2005gj - Trundle et
al. 2008).  Only lower-mass O-type stars can experience a red supergiant
phase just before or after the LBV phase (Szeifert et al. 1996; Smith et al.
2004), their evolutionary sequence being O $\rightarrow$ LBV/RSG
$\rightarrow$ WN(H-poor) $\rightarrow$ SNIb (where RSG=Red Supergiant). 
This is supported by the simultaneous presence of LBVs and RSGs observed in
Westerlund~1, the most massive young open cluster of our Galaxy, with
1$\times$10$^5$~M$_\odot$ total mass and $\leq$25~M$_\odot$ turn-off mass
(Clark et al. 2005). To the best of our knowledge, this simultaneity is not
known to occur elsewhere.

The large amount of mass lost during the LBV phase is the critical stage
that a very massive star must pass through before becoming a WR star.  Mass
ejection via major outbursts is far more efficient than that via steady
winds, as most notably undergone by $\eta$~Car during the 19th century when
the star shed several M$_\odot$ in less than a decade (Smith \& Owocki
2006). The mass lost by the central LBV star frequently gives rise to
circumstellar nebulae (e.g. Langer et al. 1999, Weis 2003) that turn out to
be enriched in processed material, such as nitrogen and helium, coming from
the stellar interior.

During the quiescent phase, the optical spectra of LBVs are characterised by
effective temperatures ranging from 12\,000 to 30\,000~K (B-spectral types)
and strong emission lines of hydrogen, HeI, FeII, CaII and other singly
ionised metals, often with P-Cygni profiles when observed at sufficiently
high resolution (cf. Kenyon \& Gallagher 1985). The mass loss rate in
quiescence is smaller and the wind faster. While not well-understood an LBV
occasionally passes through an 'outburst' phase, during which the luminosity
remains the same as in quiescence, while in tandem the mass loss rate
increases and the wind velocity decreases. This pushes outward the
pseudo-photosphere, with a consequent increase in the effective radius and
decrease in the surface temperature (whose Wien peak moves from the
ultraviolet toward the optical region). At outburst maximum, the effective
temperature is $\sim$7500-8500~K regardless of the LBV luminosity and
temperature in quiescence, and the spectrum turns to that of A-type
supergiants, while [FeII], HeI emission lines, and P-Cygni absorptions
weaken or disappear.  Wolf (1989; see Vink 2008 for an update) pointed out
the existence of an amplitude-luminosity relation for LBV eruptions and
called {\em S Doradus instability strip} the region of the HR diagram
occupied by LBVs in quiescence. Stothers \& Chin (1995) suggested the
existence of a period-luminosity relation of the form $M_{\rm bol} =
-12.5(\pm 0.5) + 2.4(\pm 0.5) \log P$ where P, expressed in yrs, is the
typical interval from one eruption to the next.

\section{RAVE Spectra of the LBVs}

  \begin{figure*}
     \centering
     \includegraphics[height=18.0cm,angle=270]{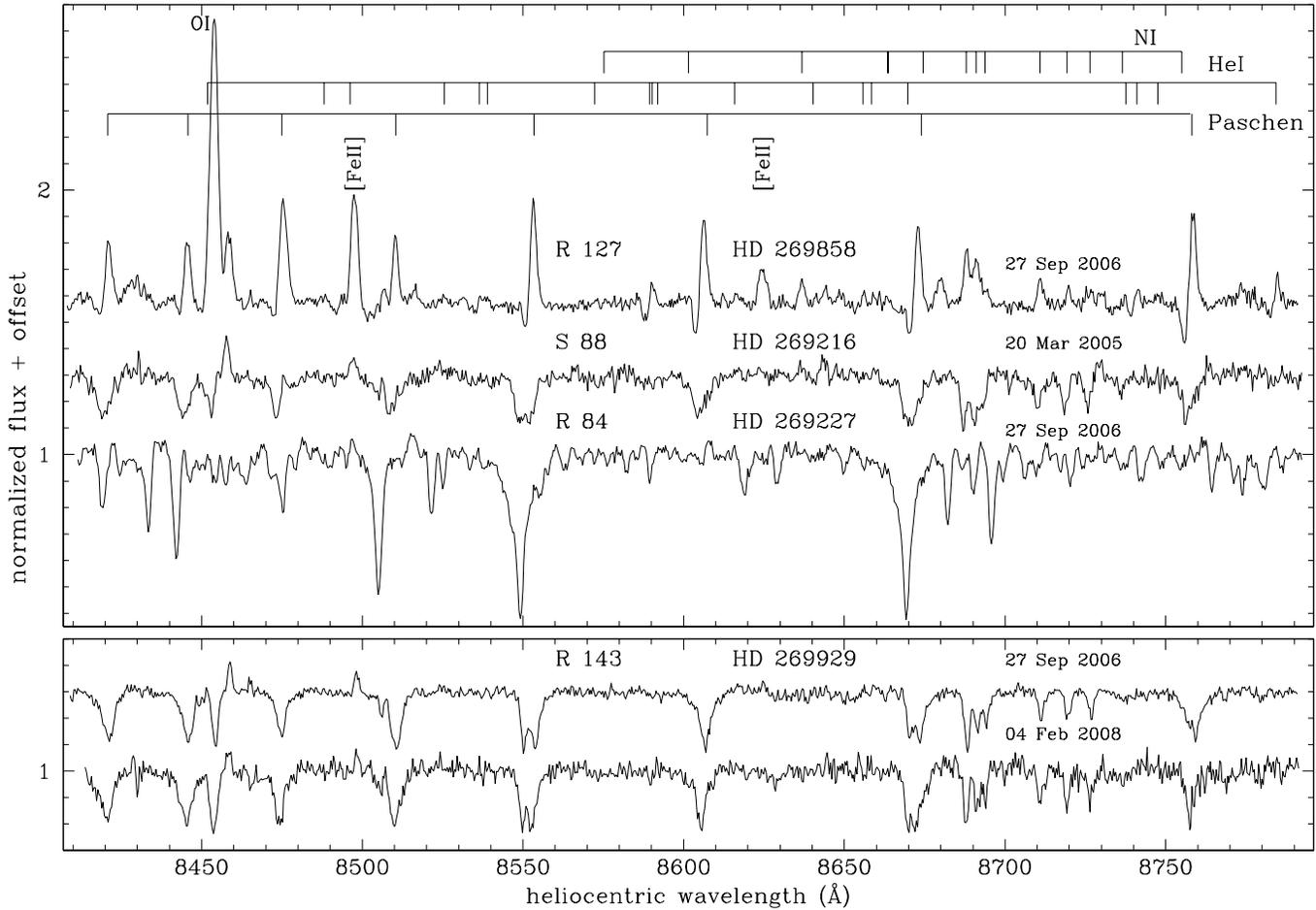}
     \caption{RAVE spectra of R~127, S~88, R~84, and R~143.}
     \label{fig3}
  \end{figure*}

The spectra that RAVE collected for the LBVs in the LMC are presented in
Figs.~1-3, where the ordinate scale is kept constant for an easier
inter-comparison. Differences in S/N among the multi-epoch spectra of the
same star are mainly due to differences in the throughputs of different
fibers and in the sky conditions.  The radial velocities derived from the
absorption and emission lines are given in Table~1 (for comparison, the
heliocentric barycentric velocity of the LMC is $+$278~km/s, Mateo 1998,
although significantly variable across the face of the galaxy).
Table~1 also lists the $V$-magnitude at the time of the RAVE observation as
estimated by consulting various databases of variable star amateur observers
(AAVSO, VSNET, VSOLJ) and the ASAS Survey database (Pojmanski 2002).

These observations indicate that the RAVE spectra are well suited to detect
and study peculiar stars, and LBVs in particular, confirming early
investigations in this wavelength interval by Munari (1998, 2002, 2003). The
peculiarities stands out well in terms of absorption features, wind outflow
signatures, and emission lines, and will now be considered in turn.

\subsection{Emission lines}

The ionised circumstellar gas is easily revealed by the rich emission line
spectrum observable over the RAVE wavelength regime. The principal ions
contributing emission lines at these wavelengths are HI, HeI, CaII, NI, CII,
CIII, OI, SI, [FeII], FeII, FeIII, [ClII], and [ClIII]. The emission lines
identified and measured in the LBV RAVE spectra of Figs.~1-3 are listed in
Table~2. They all belong either to OI, NI, [FeII], HeI, CaII or to the
Paschen series of hydrogen. 

The emission lines in RAVE spectra of LBVs are fairly ``sharp''. The expansion
velocity derived from the width of the profiles (corrected for the
FWHM(PSF)=40~km/sec instrumental resolution) is on average 25~km/sec, which
is the same as obtained from observations at more classical optical
wavelengths (Weis 2003). The only exception is R~99 (cf. Fig.~2), whose
expansion velocity from Paschen lines in RAVE spectra is $\sim$90~km/s,
comparable to the Balmer line-based results of Nota et al.  (1996). 

The intensity of the OI~8446 emission line can be boosted by fluorescence
pumping from absorption of hydrogen Lyman-$\beta$ photons by OI in its
ground state, as first pointed out by Bowen (1947). For the Lyman-$\beta$
fluorescence to be effective, the optical depth in H$\alpha$ should be
large, owing to the population of the $n=2$ level by trapped Lyman-$\alpha$
photons. Such a fluorescence pumping is at work in R~127 (cf. Fig.~3), but
not in R~99 (cf. Fig.~2), as derived by comparing the intensity of OI~8446
emission with that of the Paschen series of hydrogen, following Strittmatter
et al. (1977) analytical modelling of the pumping mechanism.

\subsection{Wind outflows}

Significant mass loss from a central star is revealed in our LBV spectra by
pronounced P-Cygni profiles. A selection of such profiles is presented in
Fig.~4. The line showing a wind-driven P-Cygni profile most consistently is
that of OI~8446~\AA.

In a P-Cygni profile, the emission component traces the systemic velocity
(eg. Castor \& Lamers 1979), so that the blueward displacement of the
absorption component relates to the velocity of the wind at the position in
the expanding medium where the given line forms (Lamers \& Cassinelli 1999).
The RAVE spectra in Fig.~4 show that the wind velocity from OI lines is slow
and quite similar in each of the surveyed LBVs, ranging from 130 to 210
km/sec. These velocities closely match those derived from high resolution
optical and ultraviolet spectra (eg. Leitherer et al. 1992; Crowther \&
Willis 1994; Garcia-Lario et al. 1998). The RAVE spectra demonstrate that
the velocity, intensity, and shape of the wind absorption component is
highly variable for any given object, as illustrated in Fig.~4 by the OI
profiles of S~Dor and R~71 at three distinct epochs. For R~71, the wind
absorption emerged only in the outburst spectrum of 4 February 2008,
contrary to the normal behaviour of LBVs.  S~Dor showed significant
variability of the OI absorption component during a period of constant
brightness. The S~Dor spectrum of 27 September 2006 shows two distinct wind
components, at 140 and 210 km/sec. Wind absorption profiles characterised by
multi-components (variable with time) have been previously observed in
high-resolution optical spectra of R~127 (Stahl et al. 1983) and AG~Car
(Stahl et al. 2001), extending over the same range of velocities detected in
the RAVE spectra of S~Dor.

The high variability of the absorption components of P-Cygni profiles
reflects the complexity of wind condition and structure in massive and LBV
stars. The winds of LBVs are both aspherical and inhomogeneous, as revealed
by the highly variable polarisation measured by Davies et al. (2005). The
amount of mass loss (in M$_\odot$~yr$^{-1}$) in such stars relates to both
luminosity (in L$_\odot$) and effective temperatures (K) as $\log \dot{M} =
1.738 \log L - 1.352 \log T_{\rm eff} - 9.547$, following the calibration of
Lamers \& Leitherer (1993). The wind velocity field is usually assumed to
follow a $\beta$-type velocity law $v(r) = v_\infty (1 - R_\star / r)^\beta$
(cf. Schmutz et al. 1991). The smaller the $\beta$, the shorter the distance
the wind needs to travel before reaching the terminal velocity. Guo \& Li
(2007) found that 0.5$\leq$$\beta$$\leq$0.7 fits the observations of LBVs in
quiescence, while $\beta$$\geq$1.5 is required for bright states. Thus, in
quiescent LBVs, $v_\infty$ is reached within a very few stellar radii,
beyond which the gas density declines as $\rho(r) = \rho(r_\circ) [r_\circ /
r]^2$, while during bright states $v_\infty$ is reached only at a much
greater distance.

A cursory inspection of the spectra of S~Dor in Fig.~1 either (i) could
suggest the presence of weak and variable emission cores within Paschen
absorption lines, or (ii) could indicate that CaII emission lines appear in
emission {\em within} the adjacent Paschen 13, 15, and 16 lines, with no
associated true P-Cygni absorption component. That this is not the case is
illustrated by Fig.~5, where a ``zoomed'' view is provided for one of the
S~Dor spectra in Fig.~1. Figure~5 clearly illustrates how the the presence
of weak emission cores within Paschen absorption lines is actually due to
the interplay with nearby NI absorption lines, and how CaII lines do really
present a P-Cygni absorption component not associated with the wings of
adjacent Paschen lines.

  \begin{figure}
     \centering
     \includegraphics[width=8.8cm]{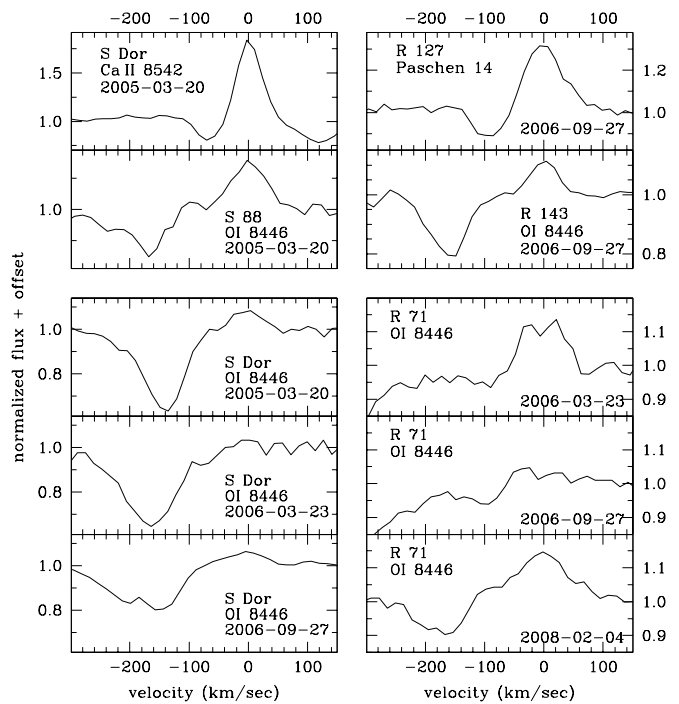}
     \caption{Sample of P-Cygni profiles of LBVs from RAVE spectra.}
     \label{fig4}
  \end{figure}
 
\subsection{Absorption spectrum and chemical abundances}

The classification of the absorption spectrum of B-A-F stars at the RAVE
wavelengths is straightforward, as illustrated by the observed MK spectral
atlases of Munari \& Tomasella (1999), Cenarro et al. (2001), le Borgne et
al. (2003), Marrese et al. (2003), and Andrillat et al. (1995), as well as
the LTE synthetic spectral atlas by Castelli \& Munari (2001) and the atlas
of non-LTE synthetic spectra of OB super-giants by Clark et al. (2005). The
absorption spectra of LBVs can be classified using these atlases, with the
caveat that both markedly non-solar surface abundance patterns and the
pseudo-photosphere being placed in a moving, partially ionised wind, can
confuse the picture (see Jaschek \& Jaschek 1987 for a self-consistent
definition of classification criteria attributes). Nevertheless, the
distinction between quiescent (B-type absorption spectra) and outburst
states (A-type absorption spectra), as well as the signatures of type Ia
luminosity class are straightforward to recognise.

Significant mass loss removes the external layers of massive stars, exposing
progressively deeper internal regions and reveals the mixing with nuclearly
processed material, which manifests as a depletion of hydrogen, oxygen, and
carbon, and an enhancement of helium and (especially) nitrogen (cf. Meynet
\& Maeder 2005). Hydrogen, helium, nitrogen, and oxygen possess strong lines
within the RAVE wavelength interval, at the effective temperatures and
gravities characterizing LBVs. Actually, multiplets 1 and 8 over the RAVE
range are the strongest NI features observable over the whole optical range
(cf. Jaschek \& Jaschek 1995). The hydrogen deficiency and nitrogen
enrichment is quite obvious in the RAVE spectra of LBVs presented in this
paper.  There is no temperature, luminosity, and metallicity combination,
for which a normal stellar atmosphere with a solar abundance pattern can
display NI lines stronger than Paschen ones as, for example, seen in the
S~Dor spectra of Fig.~1. This is both true for LTE and non-LTE synthetic
spectra (Castelli \& Munari 2001; Clark et al. 2005); for this to occur, it
would require, at the same time, {\em both} an enhancement in nitrogen and
depletion in hydrogen.

  \begin{table}
    \caption{Equivalent width of the emission lines identified and measured
             in the RAVE spectra of LBVs in the LMC.}
     \centering
     \includegraphics[width=6.9cm]{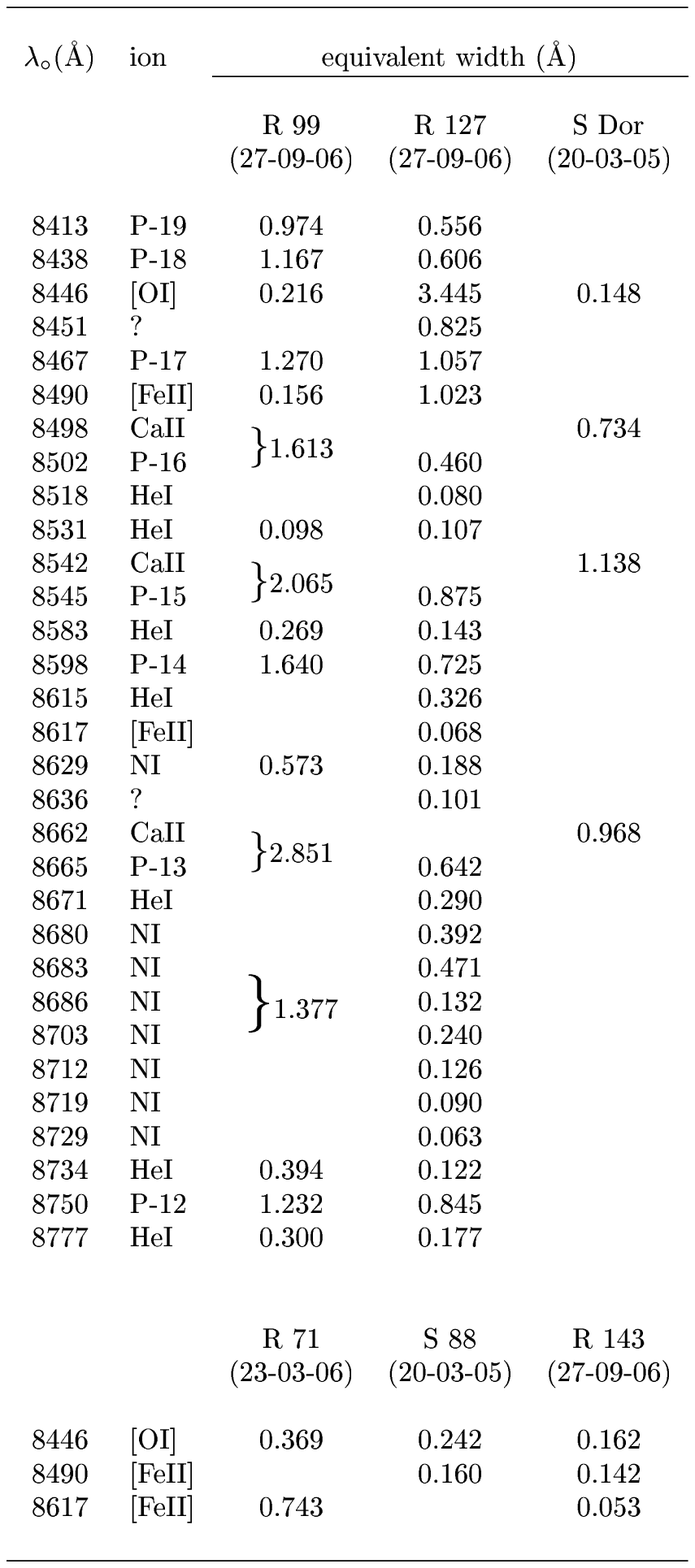}
     \label{tab2}
  \end{table}

\subsection{Radial velocities}

Table~2 presents the radial velocities of the LBVs measured from the RAVE
spectra. The accuracy of the wavelength scale of RAVE spectra is
$\sim$1.0~km/sec (Zwitter et al. 2008). The radial velocities of emission
lines for the objects with multi-epoch observations are stable within a
dispersion of 2.3 km/sec for S~Dor, 2.4 km/s for R~71, and 0.5 km/s for
R~99.  This is consistent with the long-term stability seen by Thackeray
(1974) in LBV emission line radial velocities (within a few km/s),
regardless of the phase (quiescent or outburst). Note that the observations
of R~71 in Table~2 cover both the quiescent and outburst phases.

The radial velocity of the emission lines in Table~2 are comparable to those
derived from high-resolution optical spectra at various epochs. Wasselink
(1956) derived $+$295~km/sec for S~Dor, Thackeray (1974) found $+$195~km/s
for R~71, and Weis reported $+$286~km/s for R~99. The situation for R~127 is
less clear with a low accuracy value of $+$284~km/sec claimed by Feast et
al. (1960), $+$292$\pm$9~km/s by Stahl et al. (1983), $+$267~km/s by Weis
(2003), and $+$274$\pm$1~km/s by Stahl \& Wolf (1986). The RAVE value of
$+$276.5$\pm$0.4~km/s lies within 2.5~km/sec of both the mean of these four
determinations ($+$279$\pm$5~km/sec) and the most accurate value amongst
them ($+$274$\pm$1), again consistent with a long-term constant value.
Small differences in velocity of the emission component of a P Cygni 
profile are expected when the line strength is variable.

  \begin{figure}
     \centering
     \includegraphics[width=7.8cm]{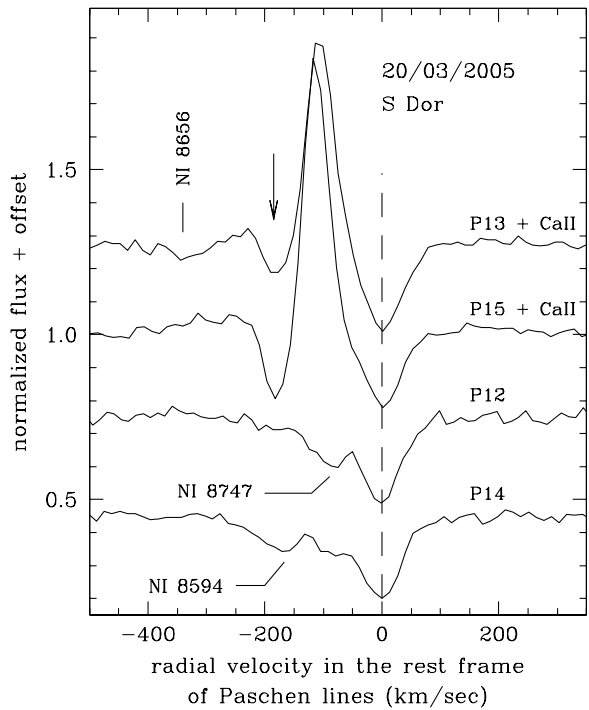}
     \caption{Expanded view of the S~Doradus spectrum of Figure~1 near the
      Paschen 12 + NI 8747, Paschen 13 + CaII 8662, and 
      Paschen 15 + CaII 8542 features. The arrow
      points to the absorption component of the CaII P-Cygni profiles.}
     \label{fig5}
  \end{figure}

On the contrary, the radial velocities of the absorption lines given in
Table~2 are highly variable from epoch-to-epoch, by several tens of km/s.
This relates to the variable positioning of the pseudo-photosphere within
the accelerating portion of the out-flowing wind, driven at least in part by
instability and inhomogeneities in the mass loss rate (Davies et al. 2005).
This 'breathing' pseudo-photosphere is also responsible for the low
amplitude, irregular photometric variability (with timescales of months to a
few years) which is persistent in LBVs both in quiescence and outburst
phases. The light-curves presented by Sharov (1975), Stahl et al. (1984),
Humphreys \& Davidson (1994), and Lamers (1995), amongst others, provide
clear examples of this restless variability. This variability can be further
appreciated by comparing the RAVE values in Table~2 with the velocities
given by Feast et al. (1960) for S~Dor ($+$213~km/sec), R~71 ($+$198), R~143
($+$263) and R~84 ($+$262).

\subsection{Photoionisation analysis}

The RAVE spectrum of R~127 provides a sufficient number of emission lines
(cf Table~2) to attempt modelling of its photoionisation structure. To this
aim, we have used the CLOUDY code (Ferland et al. 1998) to evaluate the
prevailing physical conditions in R~127 at the time of the RAVE
observations, and to assess the diagnostic potential of emission lines
observable in the RAVE wavelength regime. Even though this exercise did
provide consistent results (as outlined below), some degree of caution must
be exercised, as the number of emission lines observable over the restricted
wavelength range of RAVE is somewhat limited, and only a fraction of them is
currently treated by CLOUDY (most notably, the NI and HeI lines are
ignored). This type of modelling is best carried out by combining emission
lines from the entire optical, ultraviolet, and infrared regimes, which
provides access to a larger set of emission lines from more ions and from a
wider range of excitation conditions. Such an expanded set of input data
could also allow exploration of more complex, multi-component emission
environments, than the simple one-component geometry adopted here.

RAVE optical fibers have an aperture of 7~arcsec, which corresponds to
1.7~pc at distance of the LMC. This is large enough to include a
contribution (by unknown relative proportion) to the observed spectrum from
the central star, its wind, the extended circumstellar nebula, any parent
H\,II region, and any close optical companion (many LBVs are members of
massive and dense young stellar clusters).  Concerning R~127, it is
surrounded by a faint nebula 5~arcsec in diameter. Judging from
spatially-resolved long slit spectra presented by Weis (2003), the light
from this nebula (clearly visible on HST images obtained through an
H$\alpha$+[NII] narrow-band filter) contributes a negligible fraction of the
light compared with the unresolved central source. Similarly, there is is no
parent HII region in which R~127 is embedded, so this is also not likely to
be a problem. The adaptive optic observations by Heydary-Malayeri et al.
(2003) resolved R~127 into a dense cluster of stars, with two close optical
companions to the LBV: one is 1 arcsec away and 5 mag fainter, and the other
is 3 arcsec away and 2.7 mag fainter. The spectrum of the latter star (as
presented by Heydary-Malayeri et al. 2003) is deficient in emission lines,
and consequently, is not contaminating the LBV emission line spectrum. In
conclusion, only the ionised mass outflow from the LBV is contributing to
the emission lines observed in the RAVE spectra.

Walborn et al. (2008) have reviewed the recent photometric and spectroscopic
history of R~127, and found it entered a long-lasting outburst phase in the
early-1980s, peaking in the $V$-band at 8.9~mag during 1988-1991, and then
slowly declining toward quiescence (by the time of the RAVE observations in
2006, R~127 had reached reached this quiescent state). According to the
analysis of Guo \& Li (2007), the wind of LBVs in quiescence is accelerated
to its terminal velocity within a few stellar radii. Therefore, the vast
majority of its mass is located external to the acceleration zone and is
expanding at constant velocity. In our modelling, we thus adopted $\rho(r) =
\rho(r_\circ) [r_\circ / r]^2$ for the dependence of density on distance
from the central star. To be conservative, we also explored other density
dependencies (eg. wind accelerated to terminal velocity over significant
distances, density independent of distance from the central star), but
abandoned them ultimately due to the resulting inferior results. We assumed
Case~B nebular conditions (Osterbrock \& Ferland 2006), spherical geometry
with the photoionising star at the center, and chemical homogeneity
throughout the gas. The stellar photosphere was assumed to emit as a black
body. The chemical partition was taken to be solar for all elements except
those allowed to vary.  The initial value for the overall metallicity was
taken to be $[$M/H$]$=$-$0.5. Having no observed emission line from carbon,
and CLOUDY not treating the He and N lines recorded over the RAVE spectral
range, we assumed (and kept fixed) their abundance to the average value
found in other LBVs - i.e., a 10$\times$ enhancement for N and 2.5$\times$
for He, and a 0.2$\times$ depletion for C (cf. Crowther \& Willis 1994, and
references therein). These fixed values for He, N, and C are close to those
found by Lamers et al. (2001) for the external, spatially resolved, nebula
around R~127.
 
The emission lines that we used in the modelling, along with their observed
and computed flux ratios, are shown in Table~3. The difference between
computed and observed fluxes is $\sim$10\%. This can be regarded as an
excellent match, reflecting the limitations/capabilities of such
photoionisation modelling, regardless of the object under analysis (eg.
novae, planetary nebulae, galactic nuclei, etc.)

  \begin{table}
    \caption{Comparison between computed and observed emission lines.
             Intensities are scaled to Paschen 14.}
     \centering
     \includegraphics[width=5.0cm]{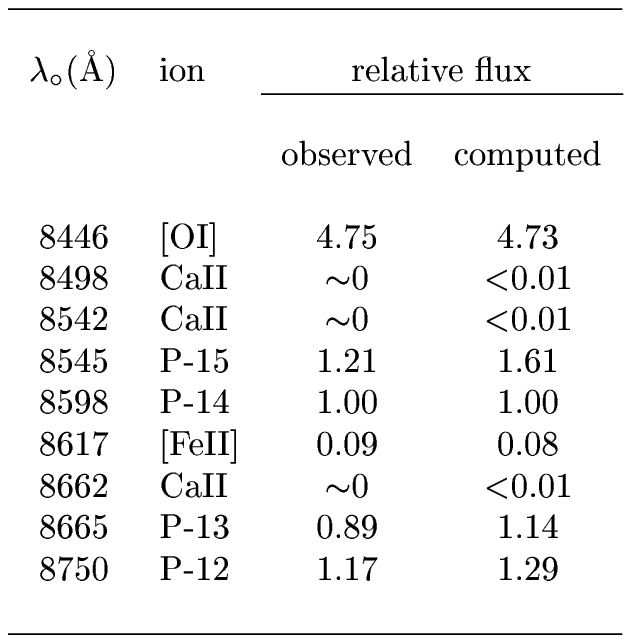}
     \label{tab3}
  \end{table}

The output parameters of the modelling are given in Table~4, where $\xi$ is
the filling factor (the ratio of filled-to-total volume in the ejecta). The
derived temperature, radius, and luminosity are consistent with other
modelling attempts in the literature, despite being based on completely
different sets of input data. For example, Stahl et al. (1983) derived from
optical and IUE high resolution observations of R~127 (during the quiescence
phase of the early-1980s), $T_{\rm eff}$=16\,000~K, $R$=150~R$_\odot$, and
$M_{\rm bol}$=$-$10.6. The hydrogen and oxygen under-abundances we have
obtained are similar to those found in other LBVs (Lennon et al. 1994;
Crowther \& Willis 1994; Venn 1997; Smith 1997), in R~127 itself (Lamers et
al. 2001), and expected by theoretical evolution models of massive stars
that include rotation (Meynet \& Maeder 2005).

The results in Table~4 indicate that the emission lines of R~127 originate
in an ionised shell detached from the central star, extending from 19 to 27
stellar radii. The mass within this ionised shell is:
\begin{equation} 
M_{\rm shell} = \frac{\xi}{X}\int_{r_{in}}^{r_{out}} 4 \pi r^2 \rho^H (r) 
dr = 1.33 \times 10^{-3}  ~~{\rm M}_\odot
\end{equation}
where X is the hydrogen mass fraction, and $\rho^H(r) = \rho^H_{\rm in}
(r_{\rm in} / r)^2$. 

The existence and meaning of such a detached shell deserves comment,
as fits squarely into the coherent picture emerging as a result of the full
suite of data associated with R~127.

Weis (2003) and Smith et al. (1998) determined from high resolution spectra
that the bulk velocity of the mass outflow during the recent outburst was
$\sim$10~km/sec. Material expanding with this velocity should have left
R~127 during 2000, in order to have reached (by the time of our RAVE
observations) a distance from the central star equal to that of the inner
border of the ionised shell in our model. It is interesting to note that the
lightcurve of Walborn et al. (2008) shows R~127 started a rapid decline in
brightness in 1999 that ended in 2000 when the star reached its ``stable''
quiescent brightness. Thus, we argue that the end of the outburst in 2000
brought with it also the cessation of an enhanced mass loss phase, producing
a cavity around the star that was only tenuously filled by the much lighter
wind characterizing the subsequent quiescence conditions.

The material at the external ionisation boundary of the shell (17.6~AU),
expanding at 10~km/sec, should have left R~127 2.5~yr before the material 
at the inner edge. The mass loss rate necessary to produce the observed shell
is therefore 
\begin{equation} 
\dot{M} = \frac{M_{\rm shell}}{\Delta t} = 5 \times 10^{-4} ~~{\rm M}_\odot/{\rm yr}
\end{equation}
This is a magnitude larger than the mass loss rate derived by Stahl et al.
(1983) for R~127 before the outburst. A large increase in the mass loss is a
characteristic feature of LBVs during outburst phases (Humphreys \& Davidson
1994). The material lost earlier in the outburst had, by the time of the
RAVE observations, moved exterior to the ionisation boundary, where the
irradiation from the central star is no longer effective in ionising the
gas, and therefore it was no longer contributing to the emission line
spectrum.

  \begin{table}
     \caption{Basic parameters for the central star and ionised circumstellar
      material derived by photoionisation modelling of the emission line spectrum of
      R~127.}
     \centering
     \includegraphics[width=5.2cm]{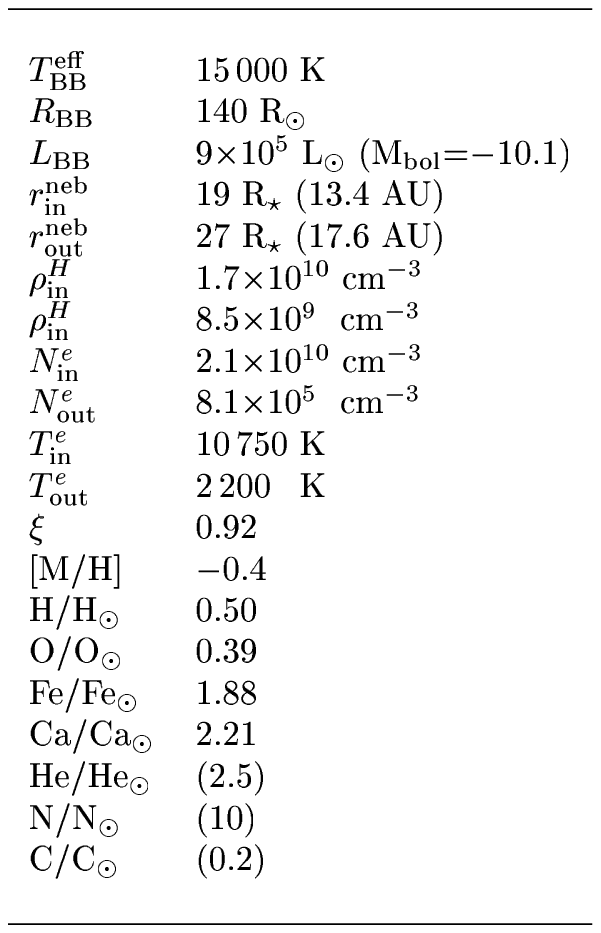}
     \label{tab4}
  \end{table}

This scenario is well matched by the spectroscopic evolution that Walborn et
al. (2008) presents for R~127. The 1999-2000 timeframe, during which the
star ejected the material comprising the ionised shell seen in the RAVE
spectra, was characterised by the disappearance of P-Cygni absorption
profiles and by the appearance of very sharp emission line profiles. These
are indicative an extended atmosphere moving outward at low velocity. Later
spectra, when the significant mass loss had finished and the detaching of
the massive shell had initiated, started showing a progressive broadening of
the emission lines and re-emergence of P-Cygni profiles characteristic of a
less-significant and faster mass loss regime.  The winds of LBVs are
line-driven, and the drop in outward velocity and large increase in mass
loss rate during outburst phases is caused by the driving effect of FeIII
lines below the sonic point as the effective temperature decreases (Vink et
al. 1999). The bifurcated wind regimes between quiescence and outburst
phases, referred to as the 'bi-stability' jump (Pauldrach \& Puls 1990), is a
defining characteristic of the S~Dor instability strip (Vink \& de Koter
2002).

\section{Remarks on Individual Objects}

\subsection{S~Doradus}

The three RAVE spectra presented in Fig.~1 were obtained while S~Dor was
stable during its outburst, at $V$$\sim$9.5~mag. Much like the Thackeray
(1965) analysis, the continuum of the RAVE spectra of S~Dor can be readily
classified as A2/3~Ia$^+$e, by comparison with the Munari \& Tomasella
(1999) atlas.

Quite interesting is the reversal in the ratio of equivalent width (REW) of
hydrogen and NI absorption lines that occurred between 20 March 2005 and 27
September 2006 spectra (cf. Fig.~1). The ratio changed from REW=0.7 to
REW=3.0. This happened at stable brightness, surface temperature, and wind
outflow velocity, while the intensity of the CaII emission components
increased by 1.6$\times$ between the two dates. It would be tempting to
interpret the reversal of REW as an indication of chemical inhomogeneities
in the nuclearly processed material progressively exposed at the stellar
surface by the on-going significant mass loss. To support this
interpretation additional and higher resolution observations are however
necessary. In fact, the observed variation in the intensity of the CaII
emission lines could suggest the presence of unresolved and variable
emission cores as a (contributing) reason for the observed changes in the
equivalent widths of nitrogen and hydrogen lines.

\subsection{R~71}

The LBV nature of R~71 was discovered by Thackeray (1974), who reported that
the star oscillates between 9.8 and 10.9 mag (see also Lamers 1995) and that
the corresponding spectral types ranged from A1eq to B2.5Iaep. During bright
optical phases, Balmer lines show deep P-Cygni profiles and the [FeII]
emission lines disappear.

The first two RAVE observations of R~71 were obtained when the star was in
quiescence at $V$$\sim$10.7, while the third was during outburst, at
$V$=9.95. The corresponding spectral changes in Fig.~2 bear a striking
resemblance to the behavior described by Thackeray (1974).  The RAVE
quiescence spectra are clearly those of an early B super-giant (strong HeI
absorption lines, no OI, and broader Paschen lines) with superimposed strong
[FeII] emission lines. The outburst spectrum is that of an early A
super-giant, where the HeI lines have disappeared and CaII and OI have
appeared, along with the reinforcement of NI and the sharpening of Paschen
lines. Similar to the optical spectra, in the outburst spectrum recorded by
RAVE, the [FeII] has nearly disappeared and OI has developed a strong
P-Cygni absorption component.

\subsection{R~84}

R~84 is an enigmatic object, suggested to be a dormant LBV by Crowther et al.
(1995). Walborn (1977) classified the blue optical spectrum as OIafpe, with
many strong emission lines (including HeII 4686) and P-Cygni profiles.
Higher resolution blue optical spectroscopy by Stahl et al. (1985) show the
same characteristics. Breysacher et al. (1999, and references therein)
summarised the determinations of the physical parameters obtained by various
authors, which cluster around a temperature of 30\,000~K, a radius of
30~R$_\odot$, a luminosity 6$\times$10$^5$~L$_\odot$ and a mass loss rate
3$\times$10$^{-4}$~M$_\odot$/yr. 

The spectral appearance at longer wavelengths is however in conflict with
that at the bluest end of the optical range. TiO absorption bands were
observed shortward of 6000~\AA\ by Sanduleak \& Philip (1977) on objective
prism plates, and by Cowley \& Hutchings (1978) in slit spectra. Allen \&
Glass (1976) found no evidence of them in far red spectra, but reported that
the MgI-$b$ triplet could be present weakly in absorption and that the IR
colors implied the presence of a cool super-giant. Stahl et al. (1984) noted
that the IR colour excess was too large to result only from a late-type
companion, and suggested the presence of dust in R~84, which was confirmed
by the mid-IR observations of Glass (1984).

These apparently contradicting reports on the spectral appearance support a
binary or optical pair nature of R~84. Some photometric variability of one
or both component stars is the probable cause of their brightness ratio to
change with time. Schmutz et al. (1991) argued strongly in favor of the LBV 
and the cool star being merely along the same line of sight and not members
of a binary system. It should be noted that adaptive optic observations by
Heydari-Malayeri et al. (2003), which have a resolution limit of 0.12 arcsec
(corresponding to 6000 AU at the distance of the LMC), did not resolve the
central star. Stahl et al. (1984) found a good fit to the IUE-optical-IR
spectral energy distribution by combining an M4Ia spectrum with a B0Ia
spectrum, and with the addition of a circumstellar dust component. The
spectral type of the cool super-giant was given as $\sim$M2 by Cowley \&
Hutchings (1978).

The RAVE spectrum of R~84 in Fig.~3 presents the clearest view so far
published of the cool super-giant. It has no emission lines superimposed and
a $\chi^2$ fitting to the same Munari et al. (2005) synthetic spectral
library used in the automatic analysis of RAVE spectra suggests atmospheric
parameters of $T_{\rm eff}$=3925$\pm$155, [M/H]=$-$0.39$\pm$0.26, $\log
g$=2.87$\pm$0.91, and a heliocentric radial velocity of
254.2$\pm$1.2~km/sec. The radial velocity reported by Schnurr et al. (2008)
is 255~km/sec. The relatively large errors of the $\chi^2$ fitting are due
to the contamination from the superimposed hot super-giant spectrum, whose
main effect is to provide an overall veiling of the absorption lines. This
has a particularly adverse effect on the wings of the CaII lines (most
sensitive to $\log g$), amplified by the perturbation of the shape of the
adjacent continuum by the wide underlying Paschen absorption lines of the
hot component. The $\chi^2$ temperature would suggest a spectral type around
K7, with the metallicity being appropriate for massive stars in the LMC.

\section{Conclusions}

The primary science driver for the RAVE survey is the investigation of the
structure and evolution of the Milky Way. Nevertheless, as a natural
byproduct of the survey, the resolution and diagnostic potential of the
associated stellar spectra ensures a valuable resource for the stellar
astrophysics community.

The present paper is the first in our series to explore the performance of
RAVE in relation to the physics of peculiar stars, in particular, the
Luminous Blue Variables of the Large Magellanic Cloud. The RAVE spectra
provide a clear and comprehensive view of the LBVs, highlighting their
temporal evolution and star-by-star differences. In particular, these have:
($i$) extended the temporal baseline and wavelength domain over which the
radial velocities of emission lines have been observed to remain constant,
which reinforces the notion of an origin in a spherically symmetric
component of the circumstellar environment, dynamically de-coupled from the
central star; ($ii$) mapped the large variability of the absorption lines
(both in radial velocity and profiles) in response to the "breathing" of the
pseudo-photosphere forming in the heavy, out-flowing wind; ($iii$) documented
the high variability and multi-shell nature of the wind P-Cyg absorption
profiles; ($iv$) provided a clear recipe of the spectroscopic changes
characterising the transition of LBVs from quiescence to outburst phases and
vice-versa, in a close match to evidence gathered at more conventional
(blue) optical wavelengths; ($v$) offered the opportunity to test
quantitatively the soundness of photoionisation modelling of the rich
emission line spectrum observable over the RAVE range; ($vii$) discovered in
R~127 the presence, and quantified the physical properties of, a massive
detached ionised shell which was ejected during the 1982-2000 outburst.

\begin{acknowledgements}
The data were obtained as part of the RAVE survey using the UK Schmidt
Telescope operated by the Anglo-Australian Observatory.  
Funding for RAVE has been provided by: the Anglo-Australian Observatory; the
Astrophysical Institute Potsdam; the Australian National University; the
Australian Research Council; the French National Research Agency; the German
Research foundation; the Istituto Nazionale di Astrofisica at Padova; The
Johns Hopkins University; the W.M. Keck foundation; the Macquarie
University; the Netherlands Research School for Astronomy; the Natural
Sciences and Engineering Research Council of Canada; the Slovenian Research
Agency; the Swiss National Science Foundation; the Science \& Technology
Facilities Council of the UK; Opticon; Strasbourg Observatory; and the
Universities of Groningen, Heidelberg and Sydney. The RAVE web-site and 
associate database is accessible from {\tt www.rave-survey.org}.
\end{acknowledgements}


\begin{thebibliography}{}

\bibitem[1976]{allen} Allen, D.A., Glass, I.S. 1976, ApJ 210, 666
\bibitem[1995]{andrillat} Andrillat, Y., Jaschek, C., Jaschek, M. 1995, A\&AS 112, 475
\bibitem[2003]{leborgne} le Borgne, J.-F. et al. 2003, A\&A 402, 433
\bibitem[1999]{breysacher} Breysacher, J., Azzopardi, M., Testor, J. 1999, A\&AS 137, 117 
\bibitem[1947]{bowen} Bowen, I. 1947, PASP 59, 196
\bibitem[2001]{castelli} Castelli, F., Munari U. 2001, A\&A 366, 1003 
\bibitem[1979]{castor} Castor, J.P., Lamers, H.J.G.L.M. 1979, ApJS 39, 481
\bibitem[2005]{cenarro} Cenarro, A.J. et al. 2001, MNRAS 326, 959
\bibitem[2005]{clark} Clark, J.S. et al. 2005, A\&A 434, 949
\bibitem[1978]{cowley} Cowley, A.P., Hutchings, J.B. 1978, PASP 90, 636 
\bibitem[1994]{crowther94} Crowther, P.A., Willis, A.J. 1994, Sp.Sc. Rev. 66, 85
\bibitem[1995]{crowther95} Crowther, P.A., Hillier, D.J., Smith, L.J. 1995, A\&A 293, 172 
\bibitem[2007]{crowther07} Crowther, P.A. 2007, ARA\&A 45, 177
\bibitem[1989]{davidson} Davidson K, Moffat, A.F.J., Lamers, H.J.G.L.M. 1989,
        eds. The Physics of Luminous Blue Variables, IAU Coll. 113, Kluwer
\bibitem[2005]{davies} Davies, B., Oudmaijer, R.D., Vink, J.S. 2005, A\&A 439, 1107
\bibitem[1960]{feast} Feast, M.W. et al. 1960, MNRAS 121, 25
\bibitem[1998]{ferland} Ferland, G.J., Korista, K.T., Verner, D.A. et al. 1998, PASP 110, 761
\bibitem[1998]{garcia} Garcia-Lario P., Riera A., Manchado A. 1998, A\&A 334, 1007
\bibitem[1984]{glass} Glass, I.S. 1984, MNRAS 209, 759
\bibitem[2007]{guo} Guo, J.H., Li, Y. 2007, ApJ 659, 1563
\bibitem[2003]{heydari03} Heydari-Malayeri, M. et al. 2003, A\&A 400, 923
\bibitem[1953]{hubble} Hubble, E., Sandage, A. 1953, ApJ 118, 353
\bibitem[1994]{humphreys} Humphreys, R.M., Davidson, K. 1994, PASP 106, 1025
\bibitem[1987]{jaschek87} Jaschek, C., Jaschek, M. 1987, The Classification of Stars Stars, Cambridge Univ. Press
\bibitem[1995]{jaschek95} Jaschek, C., Jaschek, M. 1995, The Behaviour of Chemical Elements in Stars, Cambridge Univ. Press
\bibitem[1985]{kenyon} Kenyon, S.J., Gallagher, J.S.-III 1985, ApJ 290, 542
\bibitem[1993]{lamers93} Lamers, H.J.G.L.M., Leitherer, C. 1993, ApJ 412, 771
\bibitem[1995]{lamers95} Lamers, H.J.G.L.M. 1995, in Astrophysical Application of Stellar Pulsation, R.S. Stobie and P.A. Whitelock eds., ASPC 83, 176
\bibitem[1999]{lamers99} Lamers, H.J.G.L.M., Cassinelli J.P. 1999, Introduction to Stellar Winds, Cambridge University Press
\bibitem[2001]{lamers01} Lamers, H.J.G.L.M., Nota, A., Panagia, N. et al. 2001, ApJ 551, 764
\bibitem[1999]{langer} Langer, N., García-Segura, G., Mac Low, M.-M. 1999, ApJ 520, L49
\bibitem[1985]{leitherer92} Leitherer, C., Damineli Neto, A., Schmutz, W. 1992, in Nonisotropic and Variable Outflows from Stars,
              L. Drissen, C. Leitherer and A. Nota eds., ASPC 22, 1992 
\bibitem[1994]{lennon} Lennon, D.J. et al. 1994, Sp. Sci. Rev. 66, 207, 
\bibitem[2003]{marrese} Marrese, P.M., Boschi, F., Munari U. 2003, A\&A 406, 995
\bibitem[1998]{mateo} Mateo, M.L. 1998, ARA\&A 36, 435
\bibitem[2005]{meynet} Meynet, G., Maeder, A. 2005, A\&A 429, 581
\bibitem[1999]{munari98} Munari, U. 1998, in Proccedings of the Gaia Workshop, V. Straizys ed., BalA 8, 73
\bibitem[1999]{tomasella} Munari, U., Tomasella, L. 1999, A\&AS 137, 521
\bibitem[2002]{munari02} Munari, U. 2002, in Exotic Stars as Challenges to Evolution, IAU Coll. 187, C.A. Tout and W. Van Hamme eds., ASPC 279, 25
\bibitem[2003]{munari03} Munari, U. 2003, in Gaia Spectroscopy, Science and Technology, U. Munari ed., ASPC 298, 227
\bibitem[2005]{munari05} Munari, U., Sordo, R., Castelli, F., Zwitter, T. 2005, A\&A 442, 1127
\bibitem[2008]{munari08} Munari, U., Tomasella, L., Fiorucci, M. et al. 2008, A\&A 488, 969
\bibitem[1996]{nota96} Nota, A. et al. 1996, ApJS 102, 383
\bibitem[2006]{osterbrock} Osterbrock, D.E., Ferland, G.J. 2006, Astrophysics of Gaseous Nebulae and Active Galactic Nuclei, 2nd ed., Univ. Science Books
\bibitem[1990]{pauldrach} Pauldrach, A.W.A., Puls, J. 1990, A\&A 237, 409
\bibitem[2002]{pojmanski} Pojmanski, G. 2002, Acta Astron. 52,397
\bibitem[1977]{sanduleak} Sanduleak, N., Philip, A.G.D. 1977, PASP 89, 792
\bibitem[1991]{schmutz} Schmutz, W. et al. 1991, ApJ 372, 664
\bibitem[2008]{seabroke} Seabroke, G., Gilmore, G., Siebert, A. et al. 2008, MNRAS 384, 11
\bibitem[2008]{schnurr} Schnurr, O. et al. 2008, MNRAS 389, 806
\bibitem[2008]{siebert08} Siebert, A., Bienaym\'{e}, O., Binney, J. et al. 2008, MNRAS 391, 793
\bibitem[2008]{siebert09} Siebert, A. et al. 2009, AJ, to be submitted
\bibitem[1975]{sharov} Sharov, A.S. 1975, in Variable stars and stellar evolution, IAU Symp 67, Reidel, pag. 275
\bibitem[1997]{smith97} Smith, L. 1997, in Luminous Blue Variables: Massive Stars in Transition, A. Nota and H.J.G.L.M. Lamers eds., ASPC 120, 310
\bibitem[1998]{smith98} Smith, L.J. et al. 1998, ApJ 503, 278
\bibitem[2004]{smith04} Smith, N., Vink, J.S., de Koter, A. 2004, ApJ 615, 475
\bibitem[2006]{smith06} Smith, N., Owocki, S.P. 2006, ApJ 645, L45
\bibitem[2008]{smith08} Smith, M. C., Ruchti, G. R., Helmi, A. et al. 2007, MNRAS 379, 755
\bibitem[1983]{stahl83} Stahl, O. et al. 1983, A\&A 127, 49
\bibitem[1984]{stahl84} Stahl, O. et al. 1984, A\&A 140, 459
\bibitem[1985]{stahl85} Stahl, O. et al. 1985, A\&AS 61, 237
\bibitem[1986]{wolf86} Stahl, O., Wolf, B. 1986, A\&A 154, 243
\bibitem[2001]{stahl01} Stahl, O. et al. 2001, A\&A 375, 54
\bibitem[2006]{steinmetz} Steinmetz, M. et al. 2006, AJ 132, 1645
\bibitem[1995]{stothers} Stothers, R.B., Chin, C.-W. 1995, ApJ 451, L61
\bibitem[1977]{strittmatter} Strittmatter, P.A. et al. 1977, ApJ 216, 23
\bibitem[1996]{szeifert} Szeifert, Th. et al. 1996, A\&A 314, 131
\bibitem[1965]{thackeray65} Thackeray, A.D. 1965, MNRAS 129, 169
\bibitem[1974]{thackeray74} Thackeray, A.D. 1974, MNRAS 168, 221
\bibitem[2008]{trundle} Trundle, C. et al. 2008, A\&A 483, L47
\bibitem[2001]{vangenderen} van Genderen, A.M. 2001, A\&A 366, 508
\bibitem[2008]{veltz} Veltz, L., Bienaym\'{e}, O., Freeman, K. C. et al. 2008, A\&A 480, 753
\bibitem[1997]{venn} Venn, K.A. 1997, in Luminous Blue Variables: Massive Stars in Transition, A. Nota and H.J.G.L.M. Lamers eds., ASPC 120, 95
\bibitem[1999]{vink99} Vink, J.S., de Koter, A., Lamers, H.J.G.L.M. 1999, A\&A 350, 181
\bibitem[2002]{vink02} Vink, J.S., de Koter, A. 2002, A\&A 393, 543
\bibitem[2008]{vink08} Vink, S.J. 2008, NewAR 52, 419 
\bibitem[1977]{walborn77} Walborn, N.R. 1977, ApJ 215, 53
\bibitem[1982]{walborn08} Walborn, N.R. et al. 2008, ApJ 683, L33
\bibitem[2003]{weis} Weis, K. 2003, A\&A 408, 205
\bibitem[1956]{wesselink} Wesselink, A.J. 1956, MNRAS 116, 3
\bibitem[1989]{wolf} Wolf, B. 1989, A\&A 217, 87
\bibitem[2008]{zwitter} Zwitter, T. et al. 2008, AJ 136, 421

\end{thebibliography}
\end{document}